\documentclass{mn2e}

\newif\ifAMStwofonts
\usepackage{graphicx}

\title{Spectro-Polarimetric search for hidden AGNs in four
       southern Ultraluminous Infrared Galaxies}
\author[C. Pernechele]
       {C. Pernechele,$^1$ S. Berta,$^2$ A. Marconi,$^3$ 
        C. Bonoli,$^1$ A. Bressan,$^1$
     \newauthor
        A. Franceschini,$^2$ J. Fritz,$^2$ E. Giro$^1$ \\
$^1$Astronomical Observatory of Padova, vicolo Osservatorio 5, 
I-35122 Italy \\
$^2$Department of Astronomy, University of Padova, vicolo 
Ossevatorio 2, I-35122 Italy \\
$^3$Astronomical Observatory of Firenze, L.go E. Fermi 5, 
I-50125 Italy }
\date{Accepted .
      Received ;
      in original form }

\pagerange{\pageref{firstpage}--\pageref{lastpage}}
\pubyear{}

\begin{document}

\maketitle

\label{firstpage}

\begin{abstract}
We report on a spectro-polarimetric analysis of four southern Ultra-Luminous
Infrared Galaxies (ULIRGs), aimed at constraining the presence of hidden broad
AGN lines.  For IRAS 19254--7245 ({\sl The Superantennae}) we find evidence for
a significant level of polarized light in the H$\alpha$ line with FWHM$>$2300
km/s.  Some degree of polarization is also detected in IRAS 20551-4250, though
with lower significance.  In the two other sources (IRAS 20100--4156 and IRAS
22491--1808) no polarized signals are detected. Although it is unclear from the
present data if the origin of polarization is due to reflected light from an
AGN or more simply to dichroic transmission by a dust slab, we find interesting
correlation between the presence of polarized components in the optical spectra
and independent evidence for AGN emissions in hard X-rays and the far-IR.
\end{abstract}


\section{Introduction}

With bolometric luminosities exceeding $10^{12}L_\odot$, Ultra-Luminous Infrared
Galaxies (ULIRGs) are the brightest sources in the local universe. Whether they
are powered by a hidden quasar, a circum-nuclear starburst, or a
combination of both, has been subject of a lively debate since their discovery.
Although observational hints on the nature of these galaxies have been 
increasing in the last decade, their physical understanding is still subject 
to uncertainties.
If mid--infrared ISO spectroscopy highlights clear cases of starburst--dominated
activity (Genzel et al. 1998), hard X--ray observations otherwise suggest that 
the presence of an obscured AGN may be more frequent and significant than 
predicted on the basis of the IR data alone.
Analyses of multi--wavelength (X--ray, optical and IR) observations currently
suggest that both AGN and starburst activity happen concomitantly in ULIRGs
(Genzel et al. 1998, Veilleux et al. 1999, Braito et al. 2002, 
Berta et al. 2002).

In the presence of a central engine enshrouded in a dusty torus, an appreciable
non--thermal polarized radiation may be detected, produced by either
electron or dust scattering. The use of polarimetry for study hidden broad line regions
in Seyfert 2 galaxies has been pionereed by Miller \& Antonucci (1983). 
Polarisation produced by scattering or absorptive dichroism may be also detected in 
the continuum of starburst galaxies spectra, and so the polarisation itself is not a 
straight indicator of an abscured AGN. The technique of searching for polarised 
light in IRAS galaxies has been well exploited recently by some authors 
(see for example Tran et al., 1999; Young et al., 1996). Up till now the only galaxies 
which show polarised broad H$\alpha$ all have seyfert or LINER classifications 
based on their optical spectra.

Genzel et al. (1998) studied the mid--IR spectral properties of a sample of 15
ULIRGs from a complete IRAS sample with $S_{60\mu m}=5.4$ Jy.
The far--IR selection allows to avoid selection biases due to dust absorption,
making easier the analysis of the relationship between AGN and
starburst activity.

To add further constraints on the physical processes happening within these
sources, we have performed a spectro-polarimetric survey of four of the Genzel et al.
IRAS sample in the southern sky.  We report here the results of the
polarimetric observations, while the full-light spectra are presented and
discussed by Berta et al (2002) and Fritz et al. (2002). 
A review of the optical and IR properties of these sources can be found in 
Duc et al. (1997).  Section 2 reports on
the observations and data analysis. Section 3 presents details about our
results, and Section 4 contains a short discussion.

\begin{table*}
\centering
\begin{minipage}{98mm}
\caption{Summary of the observations. Eight independent spectral
exposures have been obtained for each object (see text for details).}
\begin{tabular}{cccc}
\hline
\hline
Object  & Obs. Date (start) & Exp. Time (s) & Slit apert.(") \\
\hline
IRAS 19254--7245 & Nov. 1, 2000 & 8 $\times$ 750 & 4\\
IRAS 20100--4156 & Oct. 29, 2000 & 8 $\times$ 1200 & 3.2 \\
IRAS 20551--4250 & Oct. 30, 2000 &  8 $\times$ 1000 & 5.6\\
IRAS 22491--1808 & Oct. 31, 2000 & 8 $\times$ 900 & 8 \\
\hline
\end{tabular}
\end{minipage}
\end{table*}

\begin{figure*}
\includegraphics[height=0.48\textwidth,angle=0]{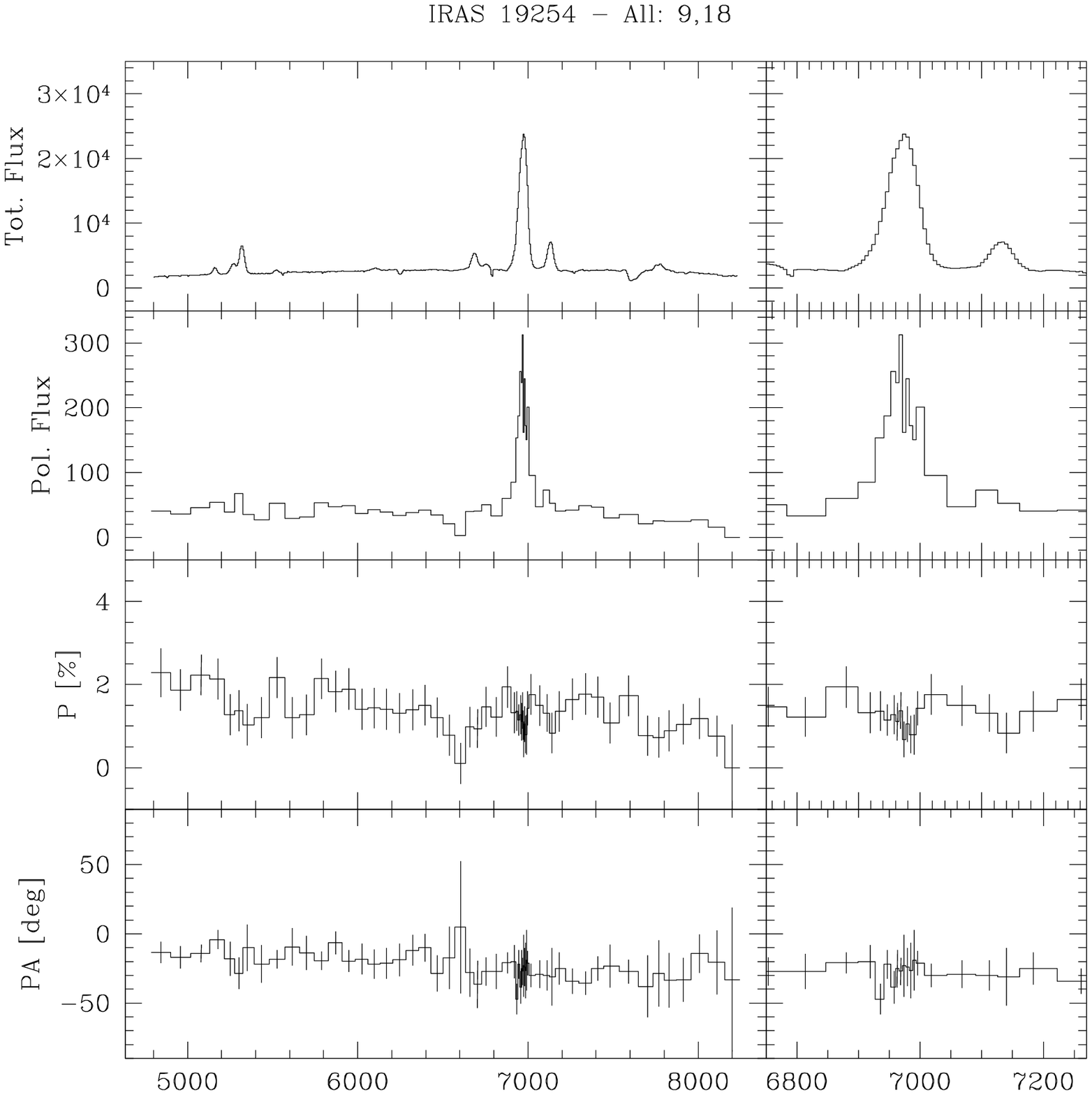}
\includegraphics[height=0.48\textwidth,angle=0]{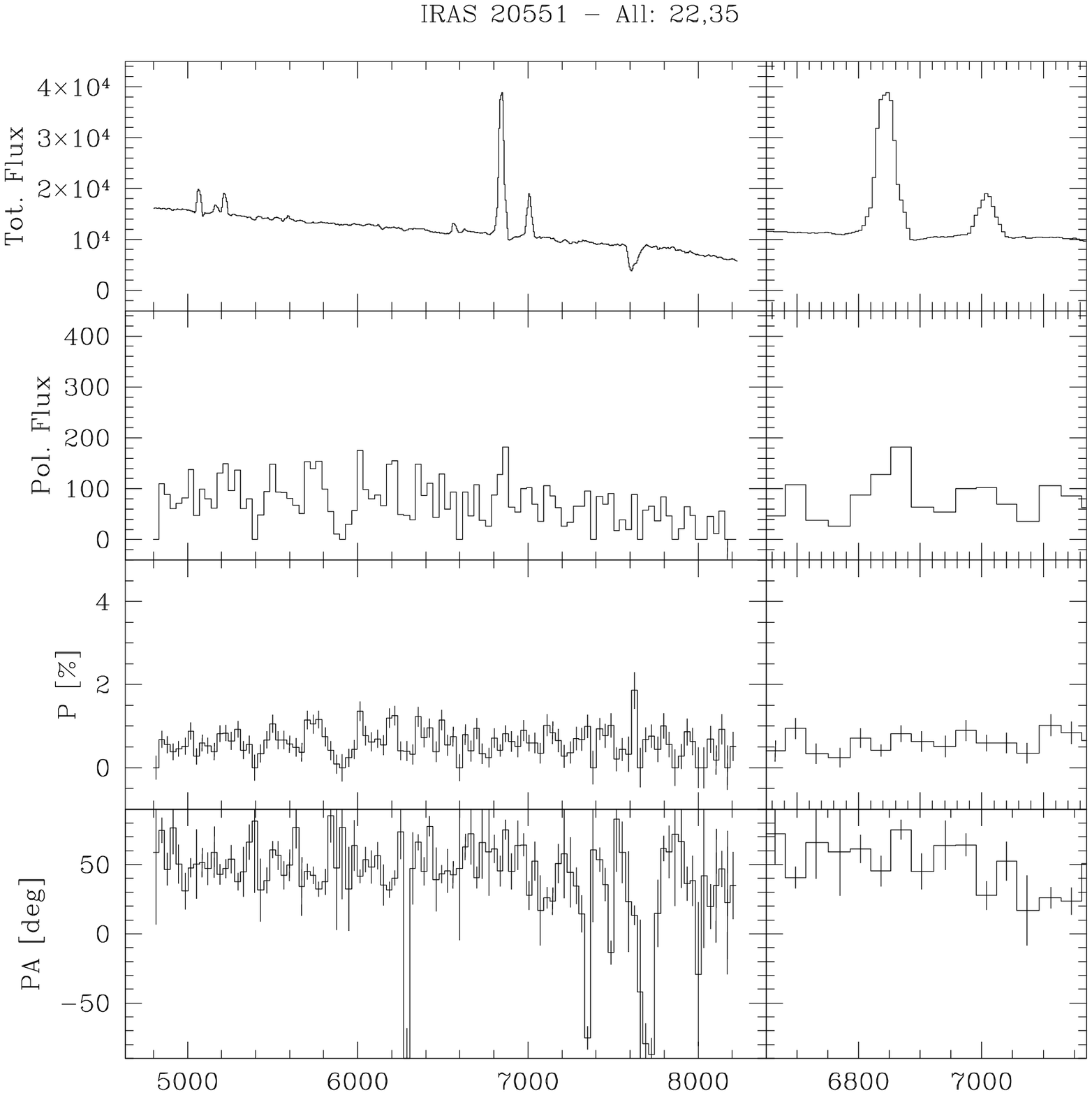}
\caption{Spectra of IRAS 19254 and IRAS 20551, the two objects for which 
evidence of polarization is found.
Each panel shows, from top to bottom (left), the total and polarized flux 
in count \AA, the polarization degree and position angle [degrees].
On the right side, zoom around the H$\alpha$ central wavelength are displayed.
Slit apertures are respectively 
4$''$ and 5.6$''$. The spectrum for IRAS 19254 refers to the southern nucleus.}
\end{figure*}

\begin{figure*}
\includegraphics[height=0.48\textwidth,angle=0]{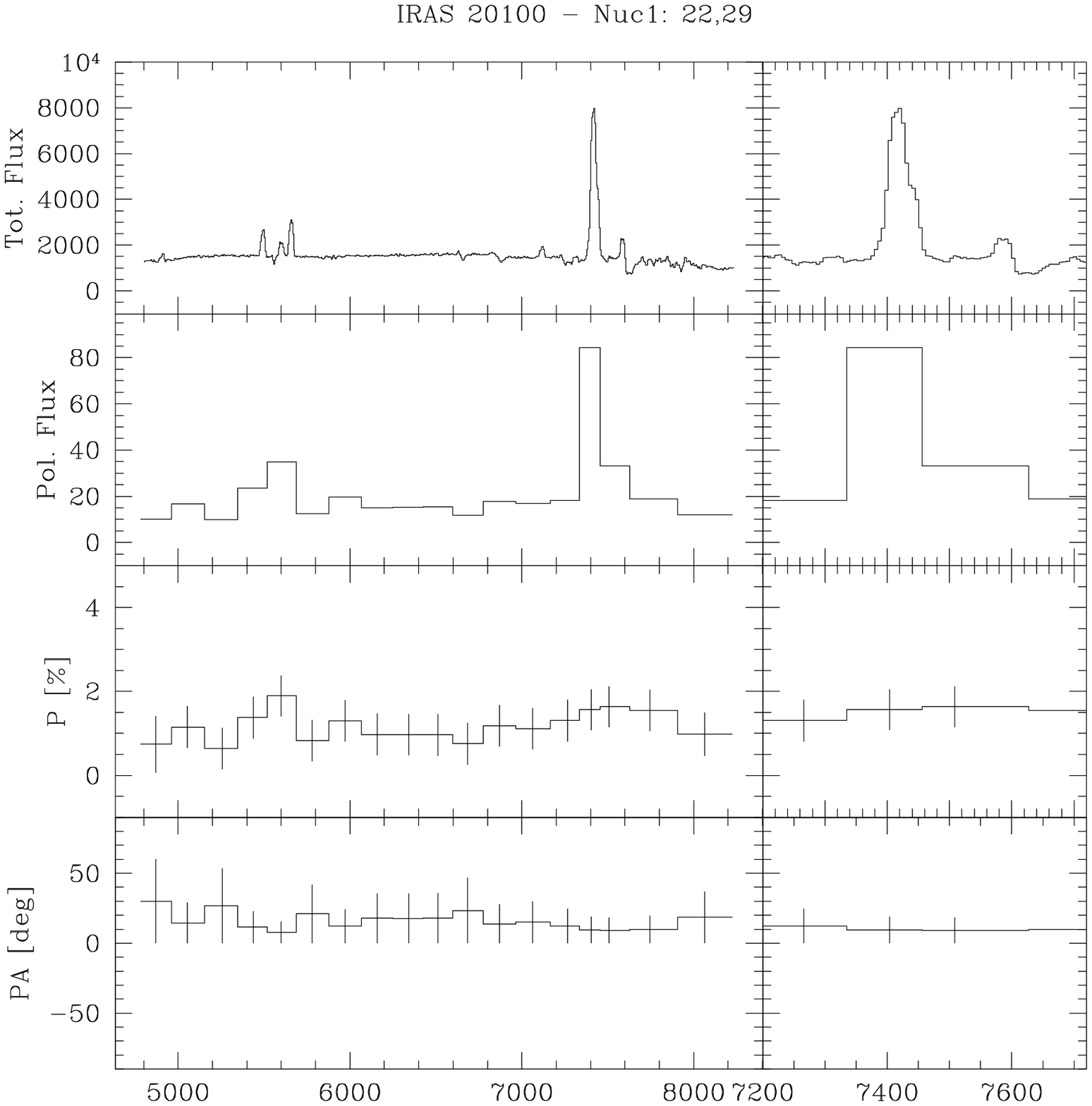}
\includegraphics[height=0.48\textwidth,angle=0]{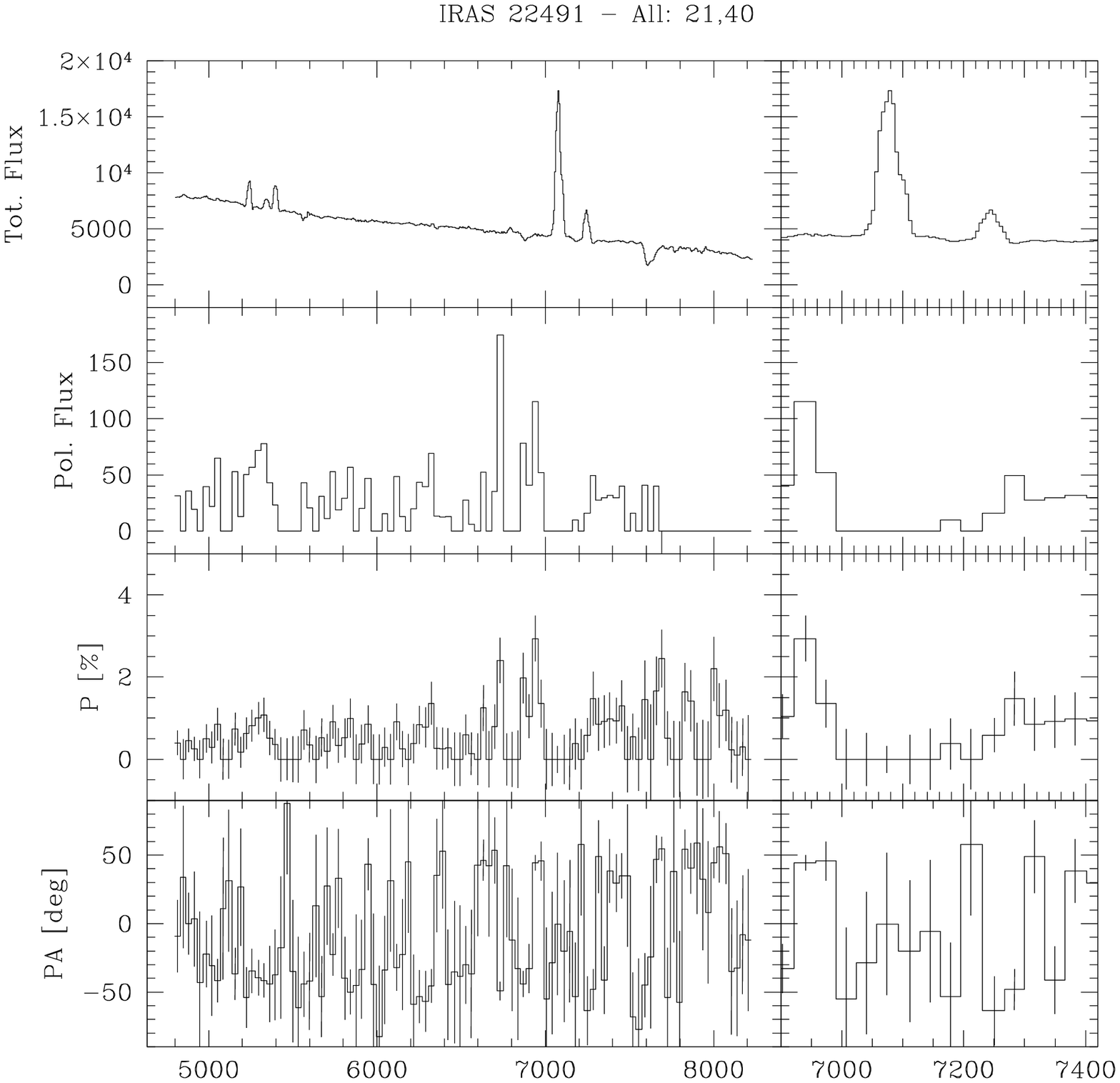}
\caption{Spectra of IRAS 20100-4156 and IRAS 22491-1808, 
in which no evidence of polarized emission is detected. 
The spectrum for IRAS 20100 is centered on the southern nucleus.
Slit apertures are respectively 3.2$''$ and 8$''$.}
\end{figure*}

\section{Observations and data reduction}

The spectro-polarimetric data have been collected with EFOSC2 at
the 3.6m ESO telescope in La Silla. Observations have been carried out during
the nights of October 29th to November 1st, 2000. Observing logs are
summarized in Table 1.
Sky was not photometric, with a seeing between 0.9 and 1.2 arcsec during the
run. We adopted the 236 line/mm grism, with an effective dispersion of
2.77 \AA/pixel at the central wavelength, operating between 3690 and 9320 \AA,
and a 1.2$\times$20 arcsec slit. This slit width corresponds to a spectral
resolution of 21.2 \AA~FWHM, or 970 Km/s in the source frame.
Position angles for the slit were chosen in order to encompass the most
interesting morphological features. In the case of galaxies with double-nuclei
(i.e. IRAS 19254-7245 and IRAS 20100-4156), the slit has been positioned along
the line joining the two nuclei.

Polarimetric observations were obtained with the combination of a
$\lambda /2$ retarder plate and a Wollastone prism. Each object was
observed with four different retarder angles (0$^\circ$, 22.5$^\circ$,
45$^\circ$ and 67.5$^\circ$), corresponding to different polarization angles
$\theta$ of the transmitted radiation. In order to check for and
eventually remove instrumental polarization, spectra of unpolarized standard
stars and light transmitted through a pinhole with a polaroid filter were
obtained.

Data reduction of the 2D spectra was performed using the standard IRAF tasks
(for details see Berta et al. 2002). Spectra were flux-calibrated by means of the
spectrophotometric standard star in NGC7293 (Oke 1990) and then corrected
for atmospheric extinction as well as for Galactic reddening. The polarimetric
reduction of the 1D extracted spectra was performed using software written
by J.R. Walsh under the MIDAS environment and IDL procedures developed by us.
Particular care has been taken in order to avoid the problem concerning 
the biased threshold always present in polarimetric measures. Debiasing has 
be done following the relation $P \sim P_{obs}[1-(\sigma_{P}/\/P_{obs})^2]^{1/2}$
where $P_{obs}$ is the observed degree of polarization, $P$ is the bias 
corrected one and $\sigma_P$ is the r.m.s. error on the polarization 
(Di Serego Alighieri, 1997).

\section{Results}

The spectra of the four observed galaxies are shown in Figures 1 and 2.
Each spectrum is obtained by identifying the trace of the main nucleus. 
On the left panels, from top to bottom, the unpolarized and
polarized spectrum, the polarization degree and direction angle are 
shown in the order, throughout the observed wavelength range. 
The panels on the right show expansions around H$\alpha$.
The plots relative to the polarization (flux, percentage and position
angle) are spectrally binned in order to have an error bar on the polarization
degree of $\pm$0.5 \% (the error bars shown in the polarization degree spectra). 
The original spectra are not rebinned. Error bars are shown also in the plots of
polarization angle.

When detected (as in the case of IRAS 19254-7245 and IRAS 20551-4250), the 
polarized lines appear to contain some structure, which however is only due to
the poor signal to noise ratio (we have checked in particular that these 
apparent
blends do not correspond to the redshifted H$\alpha$ and 
{\sc [Nii]}$\lambda$6584 lines).
Table 2 summarizes our data on the polarization degree, H$\alpha$ 
line fluxes and observed line widths.

\subsection{Note on individual sources}

{\it IRAS 19254--7245}. the {\sl Superantennae} (Colina et al., 1991), has been 
classified as a Seyfert 2 galaxy by Mirabel et al. (1991) on the basis of 
optical spectral properties.
In their polarimetric analysis, Heisler et al. (1997) and Lumsden et al. (2001)
fail in detecting scattered light and attribute the lack of polarized components 
to geometric effects: if the scattering particles lie close to the plane of 
the obscuring torus, an edge--on observer wouldn't be able to observe any 
polarized flux. In particular Lumsden et al. (2001) found only polarised
narrow emission line, and this may be explained if one consider that we were
looking at a slightly different location (Lumsden, 2002).
Moreover our total flux spectrum is essentially identical to theirs
(Lumsden, 2002).

\begin{figure}
\centering
\includegraphics[width=0.45\textwidth]{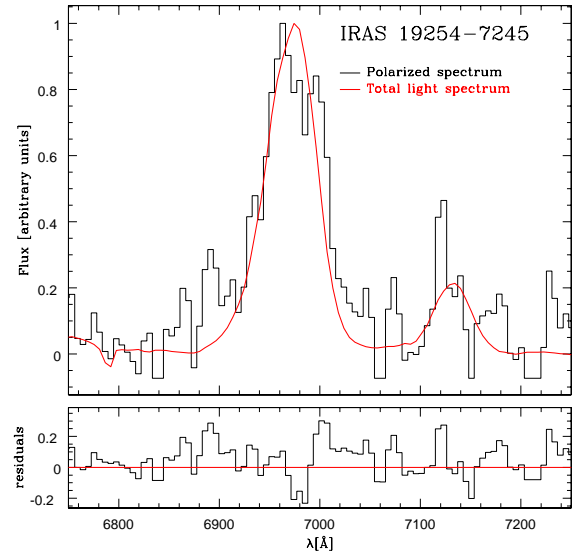}
\caption{Search for scattered polarization in IRAS
19254--7245. {\em Top}: Polarized (histogram) and total light
(thick line) spectra in the H$\alpha$ region. {\em Bottom}: difference between
polarized and full--light fluxes. The H$\alpha$ profiles in polarized and
total--light are very similar.}
\label{fig:tot_vs_pol}
\end{figure}

\begin{figure*}
\centering
\includegraphics[height=0.32\textwidth]{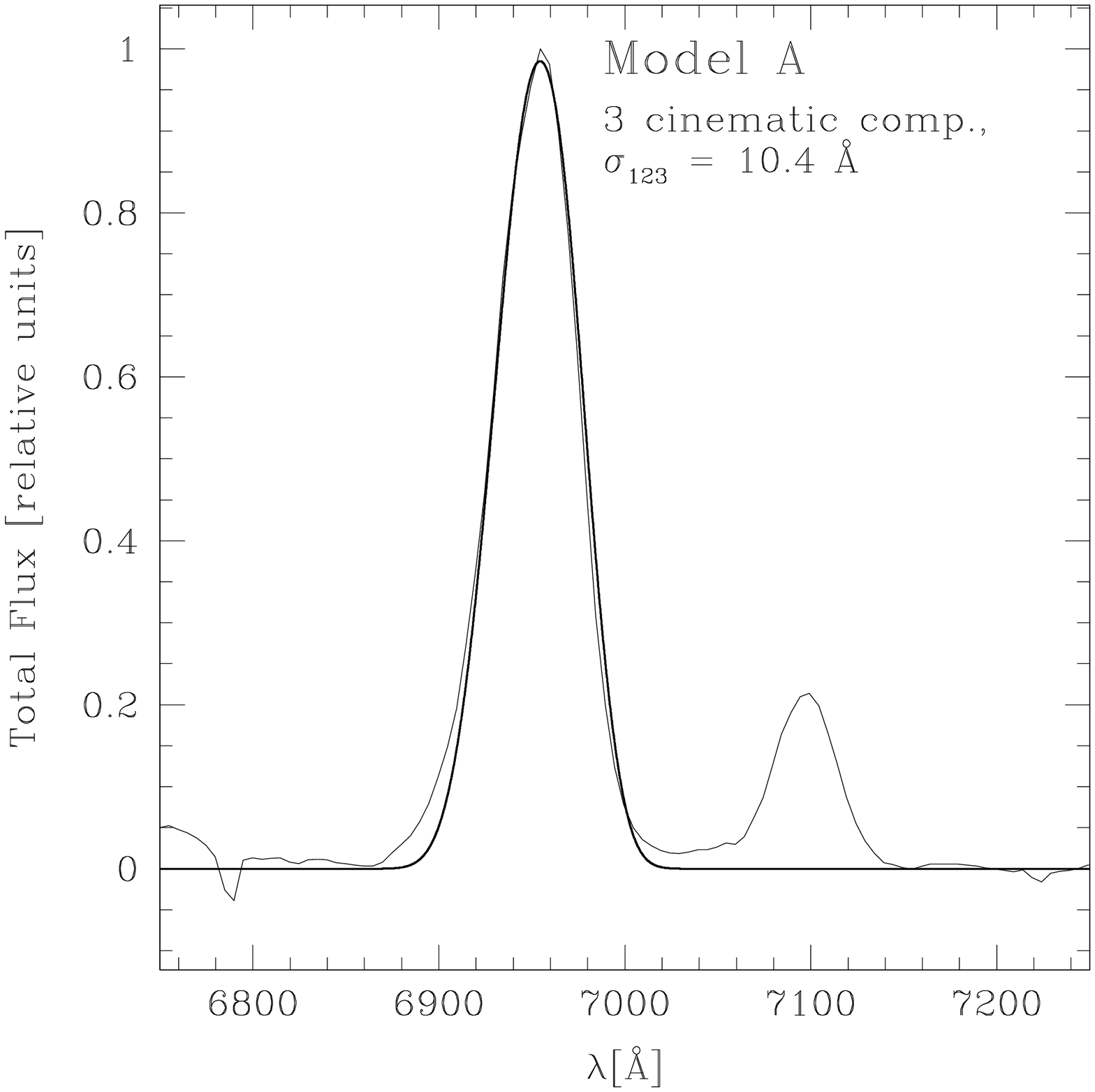}
\includegraphics[height=0.32\textwidth]{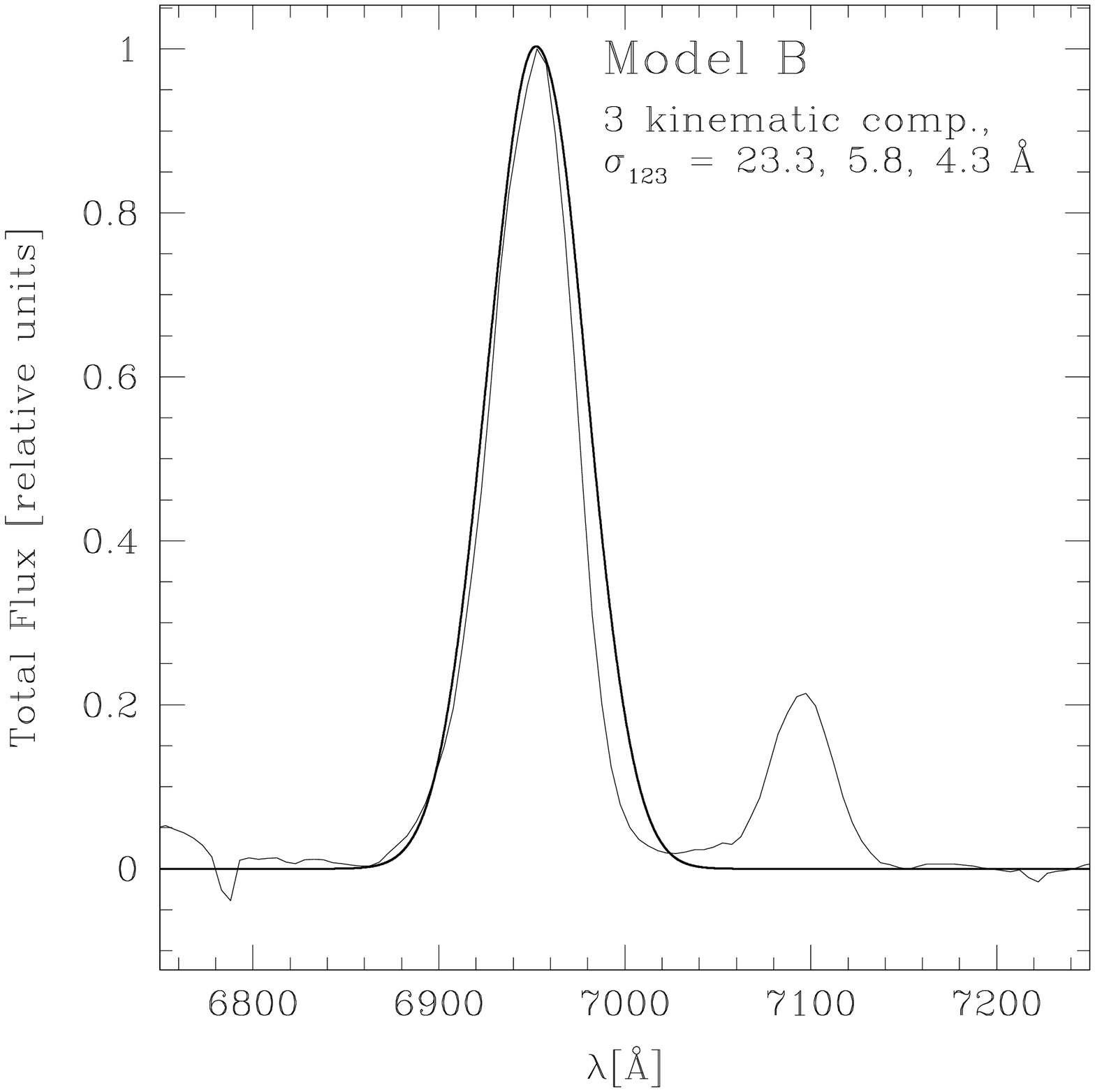}\\
\includegraphics[height=0.32\textwidth]{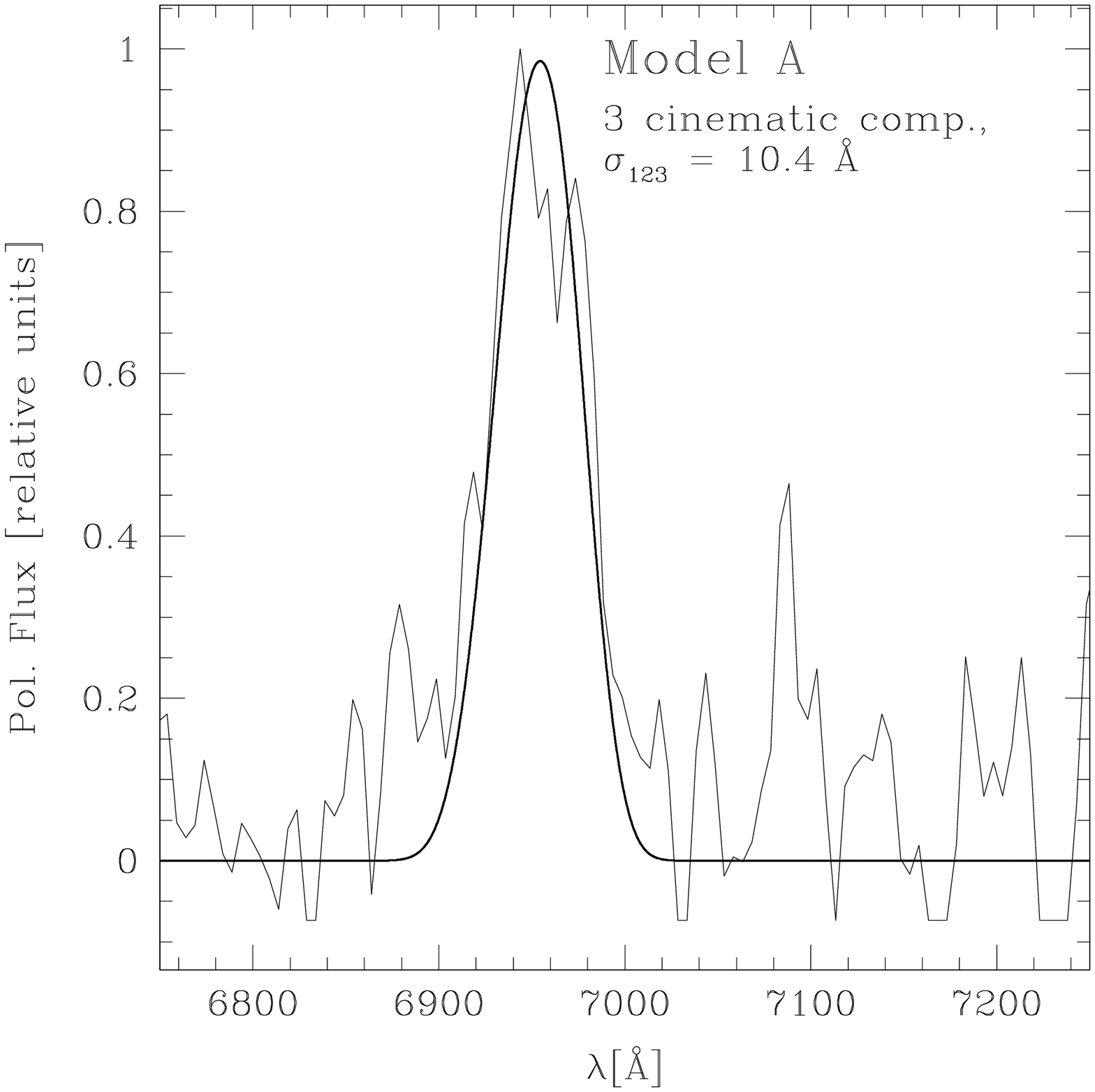}
\includegraphics[height=0.32\textwidth]{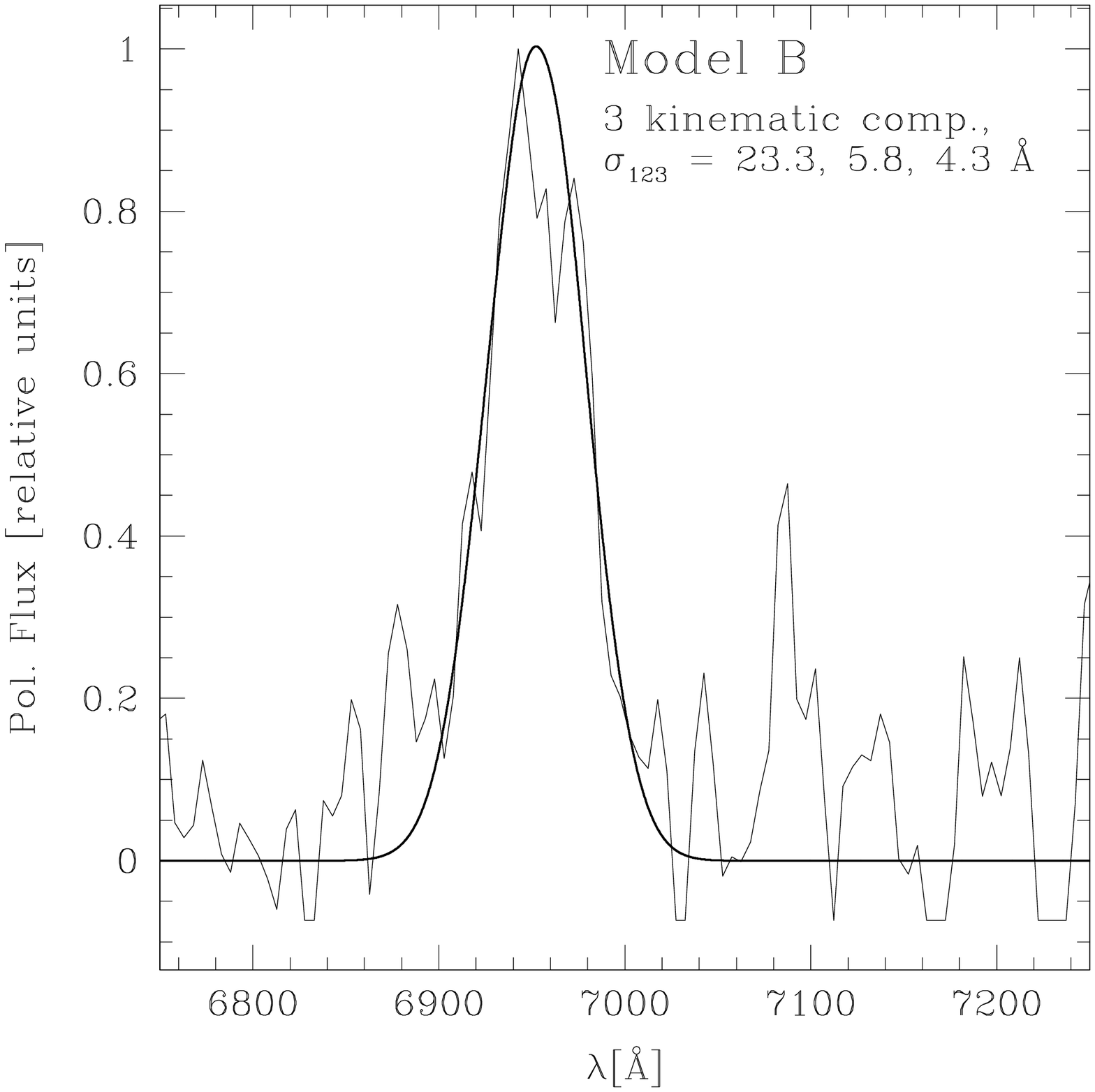}
\caption{Models (thick lines) and spectra around H$\alpha$+{\sc [Nii]}. Top panels
refer to the total flux and bottom panels to the polarized flux spectra.
``Model A'' (see text for details) and ``model B'' are shown
on the left and on right panels, respectively. Apparently,
``model A'' provides a better fit of the total flux profile (top-left panel), while the 
polarized curve is best reproduced with the ``model B'' (bottom-right), 
which has a kinematic component of $\sim 2360$ Km/s FWHM.}
\end{figure*}

The analysis of our EFOSC2 integration on the Southern Nucleus leads
to a detection of a significant polarized residual, in particular in
the spectral region around H$\alpha$ (see Figure 1, top left panel). 
The polarization degree turns out to be $\sim 1.8\pm 0.4\%$.

The width of this spectral structure in polarized light, which might be 
interpreted as the H$\alpha$+{\sc [Nii]} blend although we do not have 
the spectral resolution to resolve it, is $\sim$70 \AA, or 3500$\pm$240 km/s.
This line width is only marginally wider than that of the 
H$\alpha$+{\sc [Nii]} blend measured in the total-light spectrum by 
Berta et al. (2002, 2500 km/s).

Figure \ref{fig:tot_vs_pol} shows the comparison of the polarized to the
total--light H$\alpha$ as well as the difference between the two. There is no
clear positive residual, but the very low S/N doesn't allow to be conclusive
about the presence of a scattered component. In this figure (as in the following
figure 4) we do not have rebinned the polarized flux spectra in order to keep
the full spectral resolution around the H$_\alpha$ line.

We have attempted also to fit this polarized feature by taking into account
some information on the complex kinematic structure of the source as reported
by previously published higher-resolution work.
Following Colina et al. (1991), Vanzi et al. (2002) infer that each 
emission lines are in 
fact composed of three different kinematical components. They 
reproduce the observed optical emission lines with a synthetic profile 
with three different gaussians.
Similarly, we have fit our polarized emission structure with three different
kinematic components for the triplet {\sc [Nii]}$\lambda$6548, H$\alpha$ 
and {\sc [Nii]}$\lambda$6584 (hereafter ``model A''). The gaussian 
parameters (recession velocity $z$,
$\sigma=10.4$ \AA and percentage of flux) of Vanzi et al. (2002) were adopted.
When taking into account our lower spectral resolution $\sigma=9.0$ \AA,
the observed width of the polarized structure is not reproduced in this way.

Instead, we have found a best-fit by adding some intrinsic line-spread, i.e.
$\sigma$=23.3, 5.8, 4.3 \AA~for the three components at $z$=0.06149, 0.05858,
and 0.06414 (see Figure 4, ``model B'', Vanzi et al., 2002). Note that the sigma value
of the two latter component are lower than our instrumental spectral resolution.
In conclusion, our analysis indicate a kinematic component with an intrinsic
width of FWHM of 55 \AA, or $\sim$2360 km/s in the velocity space.

Given the complex kinematic structure of the source and our modest spectral
resolution and low signal-to-noise, we cannot be conclusive about the presence
an a broad-line AGN based on these data, particularly considering that the
full-light spectrum displays a similarly broad spectral complex.
Dychroic transmission of the H$\alpha$+[NII] complex through a
dust screen may still be consistent with the present data.

{\it IRAS 20100--4156}. Lying at $z=0.129583$, this object is extremely bright
($L_{IR}[8\div1000\,\mu m] \simeq 5\cdot10^{12}L_\odot$, 
adopting $\Omega_m=0.3$, $\Omega_\Lambda=0.7$, H$_0$=65 km/s/Mpc). 
The top left panel of Figure 2
shows the spectrum of the southern nucleus: only an upper
limit of $2\cdot 10^{-17}$ erg cm$^{-2}$ s$^{-1}$ to the polarized H$\alpha$ 
flux can be set. The northern nucleus, of comparable luminosity, does not 
show any polarization either.

{\it IRAS 20551--4250}. A detailed description of the optical
properties of ESO 286-IG19 is given by Johansson (1991).
We find very faint polarized flux in correspondance of the H$\alpha$ line
at $\sim 1\%$ level, with low significance and with a complex  structure.
The polarized feature is not spectrally resolved (see Table 2). Very similar
results on this object has been obtained by Lumsden and his collaborators 
(2002).

{\it IRAS 22491--1808}. The {\sl South America} galaxy (Carico, 1990) was 
observed in polarized light by Young et al. (1996) in the R band, but no 
polarization was detected. We confirm the lack of polarization ($<1\%$) 
and set an upper limit of
$5\cdot10^{-17}$ erg cm$^{-2}$ s$^{-1}$ to the H$\alpha$ scattered flux.

\begin{table}
\centering
\caption{H$\alpha$ polarized fluxes and FWHM estimates.
Flux values and upper limits were obtained simply integrating
the observed polarized spectrum. Upper and lower limits to line
broadenings (FWHM) have been set as described in text.}
\begin{tabular}{ccrc}
\hline
(IRAS) & degree & \multicolumn{1}{c}{H$\alpha$ flux} & H$\alpha$ FWHM \\
\hline
19254 & 1.8 & 2.5 $\cdot$ 10$^{-16}$ & 2360$\div$3500 $\pm$ 240 \\
20100 & -- & $<$2 $\cdot$ 10$^{-17}$ & -- \\
20551 &$\sim 1$ & 4.1 $\cdot$ 10$^{-16}$ & $<$1000  \\
22491 & --   & $<$5 $\cdot$ 10$^{-17}$ & -- \\
\hline
\end{tabular}
\end{table}

\section{Discussion and conclusions}

We have found very faint but significant polarization signals in two out of
four of the Ultra-Luminous IR Galaxies observed in our spectro-polarimetric
survey: these are the IRAS 19254-7245 ({\sl Superantennae}) and IRAS 20551-4250.
In the former, the polarized $H_\alpha$ line may be broadened by
$\sim$2000--3000 km/s.
The present observations tend to confirm that ULIRGs display, except in few
cases (see, for example, Young et al. 1996, Tran et al. 1999, 
Hines et al. 1999), faint signals in optical polarized light.
These results are consistent with either very weak AGN components in the 
source nuclei (otherwise dominated by starburst emission), or with a dust 
distribution covering the nuclear source and producing dichroic transmission.
The signal-to-noise ratio of the data does not allow to distinguish between
the possible polarization mechanism (e.g. electron scattering, dust scattering
or dichroism). In the case of IRAS 19254-7245 a slightly trend of polarization
degree with wavelength seems to be present (see Figure 1 left panel, third row)
and this should exclude the presence of electron scattering (that is
wavelength independent).

It is interesting to compare the present results with those based on independent
data.     All four of our sources have been observed with XMM-{\sl Newton} by 
Braito et al. (2002).
Their analysis is revealing clear evidence for AGN activity in IRAS 19254 and 20551.
In both sources, strong iron K$\alpha$ lines and large hydrogen column 
densities, indicative of hidden AGNs, are detected. 
With regard to the other two sources, the presence of non--thermal activity 
cannot be ruled out but is strongly constrained by the XMM data.

The optical-IR-millimetric spectral energy distributions of these same 
sources have also been analysed by Berta et al. (2002) and Fritz et al. (2002)
adopting combined stellar, starburst and AGN templates. AGN contributions 
to the bolometric emissions of $\sim$40\% and 18\% are indicated for 
IRAS 19254 and IRAS 20551, and of the order of few \% for the other two.
We notice interesting agreement between the results of the present 
polarimetric analysis and those based on entirely independent approaches.

\section*{Acknowledgments}

We wish to thank the technical staff of ESO 3.6m telescope at La Silla, and
particularly M. Sterzik and R. Athreya, for their efficient support.
We thanks also S.L. Lumsden for useful comments on our paper and for kindly
furnish us his data on IRAS 19254-7245 and IRAS 20551-4250.

\label{lastpage}

\end{document}

\end{document}